\documentclass[%
mathleft,%
]{an}
\usepackage{graphicx}
\usepackage{times}
\usepackage{amssymb}

\newcommand{{\aap}}{{A\&A}}% 
\newcommand\nat{{Nature}}% 
\newcommand\teff{{$T_{eff}$}}% 
\newcommand{\micron}{{$\mu$m}}% 

\begin{document}

% The following seven commands are intended for editorial usage and
% should be ignored by the author(s).
\Pagespan{1}{}% Document's page range. 
% If second parameter is left empty, the last page is computed
% automatically.
\Yearpublication{2011}%
\Yearsubmission{2011}%
\Month{1}%   
\Volume{999}%  
\Issue{92}% 
% \DOI{This.is/not.aDOI}% 

\title{The L dwarf/T dwarf transition: multiplicity, magnetic activity and mineral meteorology across the hydrogen burning limit.}

\author{Adam J.\ Burgasser \inst{1}\fnmsep\thanks{Corresponding author:
  \email{aburgasser@ucsd.edu}}
% Example for footnote, note the usage of the \texttt{fnmsep} command
% as separator between institute number and footnote mark}
}
\titlerunning{The L/T Transition}
\authorrunning{A.\ J.\ Burgasser}
\institute{
University of California San Diego, MC 0424, 9500 Gilman Drive, La Jolla, CA 92093, USA}

\received{XXXX}
\accepted{XXXX}
\publonline{XXXX}

\keywords{infrared: stars, stars: binaries: spectroscopic, stars: low-mass, brown dwarfs, techniques: spectroscopic}

\abstract{%
The transition between the L dwarf and T dwarf spectral classes is one of the most remarkable along the stellar/brown dwarf Main Sequence, separating sources with photospheres containing mineral condensate clouds from those containing methane and ammonia gases.  Unusual characteristics of this transition include a 1~$\mu$m brightening between late L and early T dwarfs observed in both parallax samples and coeval binaries; 
a spike in the multiplicity fraction;
evidence of increased photometric variability, possibly arising from patchy cloud structures;
and a delayed transition for young, planetary-mass objects.  
All of these features can be explained if this transition is governed by the ``rapid'' (nonequlibrium) rainout of clouds from the photosphere, triggered by temperature, surface gravity, metallicity and (perhaps) rotational effects.  
While the underlying mechanism of this rainout remains under debate, the transition is now being exploited to discover and precisely characterize tight ($<$ 1 AU) very low-mass binaries that can be used to test brown dwarf evolutionary and atmospheric theories, and
resolved binaries that further constrain the properties of this remarkable transition.
}
\maketitle

\section{Introduction}

The L dwarf and T dwarf spectral classes encompass what were until recently the lowest-temperature, lowest mass (VLM) stars and brown dwarfs known, with {\teff} $\approx$ 600--2200~K (see contribution by J.\ D.\ Kirkpatrick on the newly-defined Y dwarf class).  
While these classification classes are defined at different wavelengths (Kirkpatrick et al.\ 1999; Burgasser et al.\ 2006a),
they are readily distinguished by their infrared (IR) spectra, 
dominated by H$_2$O, CO, FeH, and alkali absorption for the L dwarfs
and H$_2$, H$_2$O, CH$_4$, and NH$_3$ absorption for the T dwarfs.  The IR spectral energy distributions (SEDs) of these classes are also distinct,  L dwarfs being quite red 
in near-IR colors ($J-K \sim$ 1.5--2.5) and T dwarfs being relatively blue ($J-K \lesssim$ 1).
This variation is due in large part to condensate cloud opacity, which arises as TiO, VO, Fe and other metals condense out of the atmosphere to form 10-30~{\micron} grains in the photosphere (Ackerman \& Marley 2001; Helling et al.\ 2008).  These grains scatter efficiently across the near-IR, having their greatest influence in the 1.0--1.3~{\micron} gas opacity window and giving rise to the red colors and muted H$_2$O bands in the near-IR spectral of these objects (Allard et al.\ 2001).
%T dwarf spectra are dominated by H$_2$O, CH$_4$, and NH$_3$ absorption, and have relatively blue SEDs ($J-K \lesssim$ 1) reflecting a depletion of photospheric grains and the increasing role of collision-induced H$_2$ absorption in the 2~{\micron} $K$-band region (\cite{1994ApJ...424..333S,1997A&A...324..185B}).
There are also pronounced variations in L and T dwarf SEDs that correlate with differences in surface gravity, metallicity, condensate cloud properties, and nonequilibrium chemistry 
(Looper et al.\ 2008; Burgasser et al. 2008; Cushing et al.\ 2010).

\section{Evolution across the L/T transition}

Given their distinct spectral morphologies, and the fact that all brown dwarfs
evolve through the L and T spectral sequence as they cool, the transition between these classes---defined here as spectral types L8 to T5---is key to understanding the atmospheric chemistry, thermal evolution and statistical distribution of brown dwarfs in the Galaxy.
Formally, the start of the T dwarf class is defined by the emergence of the 1.6~{\micron} CH$_4$ band, but the observed changes in SED across the L/T transition also reflects a sudden removal of condensate grain opacity from brown dwarf photospheres over a relatively narrow {\teff} range.  
The rapidity of this shift is evident in the brightening of early-type T dwarfs
around 1~{\micron}---the so-called ``J-band bump''---that is observed in both parallax samples (Tinney et al.\ 2003; Vrba et al. 2004) and between components of binaries spanning the transition (Burgasser et al.\ 2006b; Liu et al.\ 2006; Looper et al.\ 2008; see below).  
%As seen in Figure~X of KIRK12, this brightening is only the second inflection point in the $H$-band magnitude-spectral type relation, after the formation of $H_2$ at the K dwarf/M dwarf transition. [CUT?]
As opacity in this region is dominated by condensates in the L dwarfs, the observed brightening, along with the resurgence of FeH and alkali line absorption in the 1~{\micron} region (Burgasser et al.\ 2002), indicates a loss of condensates rather than obscuration from emergent CH$_4$ absorption.  
A relatively sharp transition is also suggested by the 
small change in luminosity that encompasses the transition (Golimowski et al.\ 2004),
the rarity of early-type T dwarfs in current search samples (Metchev et al.\ 2008),
and an apparent excess of binaries at the L/T transition (Burgasser 2007a).  
These lines of evidence suggest that the L/T transition may be 
a relatively short-lived phase of brown dwarf evolution, although consideration
must be taken into account for possible feedback in the thermal evolution
at this phase (Saumon \& Marley 2008).
  
  \begin{figure*}[t]
\includegraphics[height=6cm]{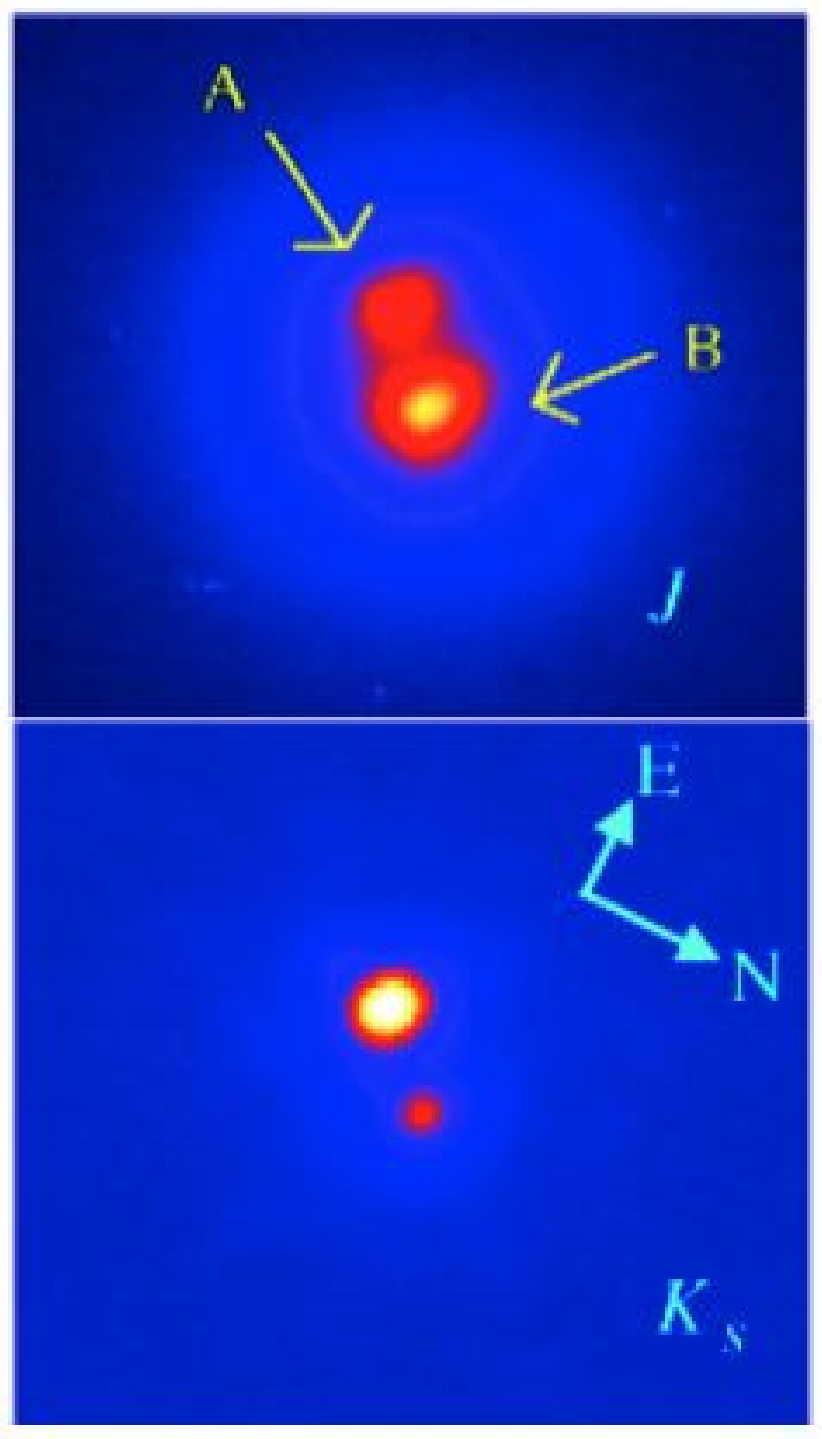}
\includegraphics[height=6cm]{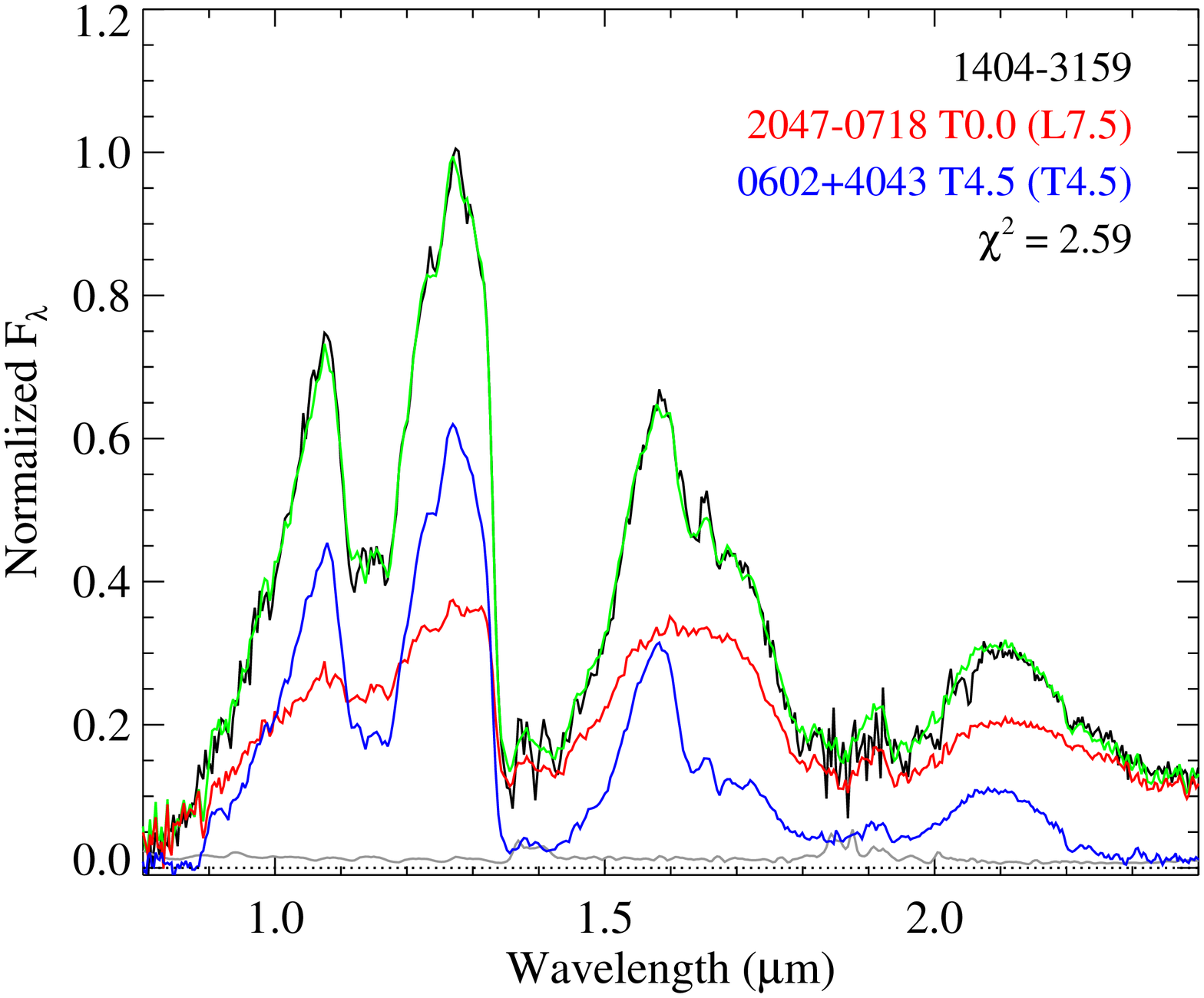}
\includegraphics[height=6cm]{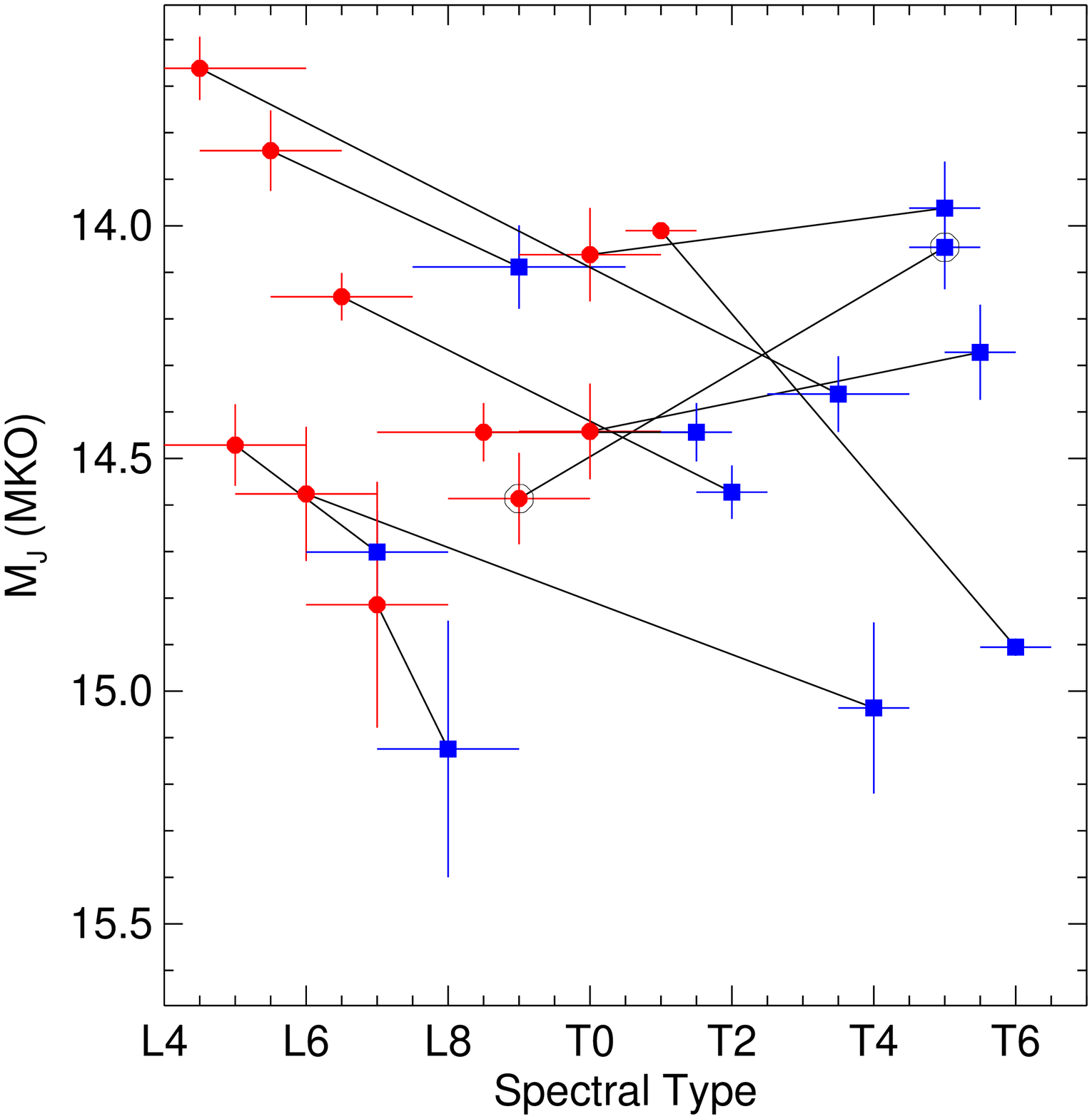}
\caption{
{\it Left}: Adaptive optics images of the L/T transition binary 2MASS~J1404-3159AB at $J$ (top) and $K_s$ (bottom).  Primary (A) and secondary (B) are labeled, and are designated based on their $K_s$-band flux (from Looper et al.\ 2008).
{\it Center}: Combined-light near-infrared spectrum of 2MASS~J1404-3159AB (black line) compared to the best fit template binary (green or medium grey line) composed of a T0 primary (red or dark grey line) and a T4.5 secondary (blue or lower light grey line), based on resolved photometry. The secondary is considerably brighter than the primary at 1.05~{\micron} and 1.27~{\micron} (from Burgasser et al. 2010). 
{\it Right:} Absolute MKO $J$-band magnitudes of resolved L/T transition binaries with measured parallaxes as a function of component type, based on the compilation of Dupuy \& Liu (2012).  Primaries are indicated by red (or light grey) circles, secondaries by blue (or dark grey) squares, and pairs are joined.  Lines that slope upward, including 2MASS~J1404-3159AB (large black circles), are flux reversal binaries.
}
\label{fig:bin1404}
\end{figure*}

Static 1D atmosphere models for brown dwarfs which parameterize condensate cloud properties (e.g., $f_{sed}$ in Ackerman \& Marley 2001) predict a much slower decline in condensate opacity with {\teff} as cloud tops sink to deeper layers. (Marley et al.\ 2002).  To reproduce the transition,
a change in cloud surface coverage or condensate grain properties must be introduced. The first possibility, inspired by the patchy tropospheric clouds of Jupiter and Saturn, predicts that holes in the cloud layer form at the L/T transition, allowing ``hotter'' light to emerge from below (Ackerman \& Marley 2001; Burgasser et al.\ 2002; Marley et al.\ 2010).  The second possibility, a global change in the cloud thickness at the L/T transition, may reflect a ``rainout'' of the upper cloud decks by dynamic perturbations (Knapp et al.\ 2004; Saumon \& Marley 2008).  Either of these cases may be driven by changes in the convective structure of the photosphere; properly tuned, both can reproduce the $J$-band brightening and changes to brown dwarf SEDs (Cushing et al.\ 2008)

A potential discriminator of these hypotheses is the possibility that cloud fragmentation could induce more frequent and intense variability at the L/T transition, particularly around 1~{\micron}, as cool clouds and hot holes rotate in and out of view.  
The recent detection of two strongly variable brown dwarfs, SIMP J0136+0933 (Artigau et al.\ 2009)
and 2MASS J2139+0220 (Radigan et al.\ 2012), both of which are early-type T dwarfs, are seen to support this idea.
However, there is currently limited theoretical work on cloud structure in rapidly rotating brown dwarfs to make a robust prediction of this effect (Schubert \& Zhang 2000; Sanchez-Lavega 2001), and variability samples are currently too sparse to robustly identify trends.
As both H$\alpha$ and radio emission have been detected in cool T dwarfs (Route \& Wolszczan 2012), magnetic spots or auroral emission cannot be ruled out as a source for this variability (Lane et al.\ 2007; Berger et al.\ 2009).

\section{Binaries and the L/T transition}

\subsection{Resolved binaries: tracing excess}

One of the key lines of evidence that condensates are removed from the photospheres of L/T transition objects is the 1~{\micron} brightening observed between ``flux reversal'' binaries that straddle this transition.  Figure~\ref{fig:bin1404} shows an example of one such source, 2MASS~J1404-3159AB (Looper et al.\ 2008), whose T4.5 secondary is $\sim$0.5~mag brighter at $J$ and $\sim$1~mag fainter at $K$ than its T0 primary.  The excess is concentrated in the 1.05~{\micron} and 1.27~{\micron} flux peaks, the regions most affected by condensate opacity in the L dwarfs.
However, not all binaries that straddle the L/T transition are flux-reversal systems; there is in fact considerable variation in the absolute brightnesses and relative flux trends among these systems, suggesting that additional secondary parameters play a role in governing when and how quickly brown dwarfs evolve across the L/T transition (see below).

\subsection{Spectral binaries: probing small separations}

\begin{figure}
\includegraphics[width=0.9\linewidth]{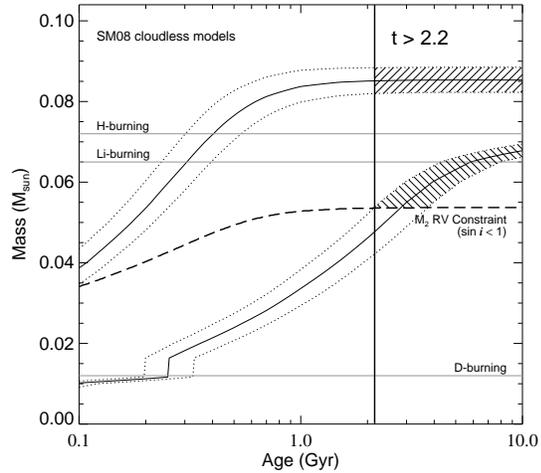}
\caption{
Age and mass constraints for the tight spectral binary 2MASS~J0320-0446AB, based on component spectral types (curved solid and dotted lines), RV orbit  (dashed line) and  evolutionary models from Saumon \& Marley (2008).  This analysis constrains the primary mass to 0.085$\pm$0.025~M$_{\odot}$ and the age of the system to $>$2~Gyr (from Burgasser \& Blake 2009).}   
\label{fig:age0320}
\end{figure}

The dramatic SED changes that occur
across the L/T transition also facilitate the identification of spectral binaries, or blended-light pairs.  One of the first L/T binaries discovered, 2MASS~J0518-2828AB, was initially recognized as a peculiar source whose near-infrared spectrum could be reproduced as a late L dwarf and mid-type T dwarf composite (Cruz et al.\ 2004).
Since then, nearly two dozen spectral binaries and binary candidates have been identified
(Burgasser 2007b; Burgasser et al.\ 2008a, 2010, 2011, 2012; Gelino \& Burgasser 2010; Kirkpatrick et al.\ 2010; Gei$\beta$ler et al.\ 2011).
Some have been resolved, allowing for robust assessments of their component SEDs, orbital characteristics and physical properties.  
%An illustrative case in point is 2MASS~J1315-2649AB, a resolved L5 + T7 spectral binary with $\Delta$K = 5~mag, separated by 0$\farcs$34 (6.6~AU).  The primary is an unusually active L dwarf, and coevality analysis of the two components reveals that this system is relatively mature ($>$1~Gyr), an apparent reversal in standard age-activity relations.  The stellar mass of the primary may explain its latent activity [\cite{2010A&A...522A..13R}].
However, spectral binaries also allow us to identify systems that are too tightly separated or too distant to be resolved by direct imaging.  This is relevant to our understanding brown dwarf formation, as the incidence of tight brown dwarf binaries ($\rho <$1~AU), and hence the overall binary fraction, remains poorly constrained (3--30\%; Basri \& Reiners 2006; Joergens 2008; Blake et al.\ 2010).  
Identification of these intrinsically faint systems through radial velocity (RV) monitoring and microlensing programs remains inefficient.  
%Nevertheless, tight binaries may be common, given the sharp cutoff in the angular separations of known VLM binaries at the limit of high-resolution imaging (Figure~\ref{fig:binfrac}). 
Yet two of the four known VLM binaries with $\rho <$0.5~AU 2MASS~J0320-0446AB (Burgasser et al.\ 2008; Blake et al.\ 2008) 
and SDSS~J0006-0852AB (Burgasser et al.\ 2012), were both identified as spectral binaries. 
%Both are M8.5+T5 pairs exhibiting RV variations of 6--8~km/s and orbital periods of 8~months and 5~months, respectively (the latter is also part of a hierarchical VLM triple system).  
By combining the component {\teff}s of these systems (from spectral binary analysis) with their system mass function (from the RV orbit), one obtains constraints on the age and orbital inclination of these systems based on evolutionary models (Fig.~\ref{fig:age0320}).  Where independent age information is available, the evolutionary models themselves can be tested.\footnote{This is the case for SDSS~J0006-0852AB, where the lack of H$\alpha$ emission in both SDSS~J0006-0852A and its wide companion LP 704-48 indicate an age $\gtrsim$8~Gyr, consistent with the $\gtrsim$4~Gyr age inferred from the analysis described above.}  Tight spectral binaries therefore complement studies of astrometric binaries in testing brown dwarf theory (Dupuy et al.\ 2009; Konopacky et al. 2010), but on considerably shorter timescales.  
%Work is ongoing to monitor unresolved spectral binaries for RV variations.
%However, unlike direct imaging programs, the initial identification of a spectral binary candidate is not dependent on that systems actual or projected separation.  This feature can in principle provide improved constraints on multiplicity statistics, by  increasing the number of systems examined for multiplicity (identified at greater distances and hence larger search volumes) and by identifying systems with very small separations ($a <$ 1~AU). Improved binary numbers has indeed come to fruition, as roughly two dozen spectral binaries and binary candidates have been identified in the past five years (REFS).  This reflects a 20\% increase in the known sample of very low mass pairs.  The SEDs of the components of these systems, inferred by decomposition of combined-light spectra with an assumed spectral type-absolute magnitude relation, indicates the same kind of variation in the degree of brightening at 1~{\micron} as observed in resolved L/T binaries.  

%As the number of confirmed spectral binaries increases, it should be possible in principle to determine the intrinsic binary fraction of VLM dwarfs independent of separation.  However, the requirement that these systems contain certain combinations of spectral types---typically late-M and L dwarfs with mid-T dwarf companions---which are mass and age dependent for brown dwarfs, results in a complex selection function.  Furthermore, 
Not all blended-light systems will turn out to be physical binaries.  A case in point is the photometrically variable T2.5 dwarf 2MASS~J2139+0220, which was identified as a spectral binary (Burgasser et al.\ 2010), but is unresolved in HST imaging (Radigan et al. 2012) and does not appear to be an RV variable (Khandrika et al.\ 2012). The spectral binary nature of this source may instead reflect the distinct SEDs emerging from cloudy and non-cloudy regions on its surface, and may suggests a method of finding other highly variable sources.

\section{Parameters governing the L/T transition}

While the spectral changes encompassing the L/T transition are driven largely by condensate depletion, what serves as the trigger? 
Temperature is clearly a factor, as analyses of color-magnitude diagrams suggest that most brown dwarfs transition from late-L to early-T over a narrow {\teff} range (100--200~K) around 1200~K (Faherty et al.\ 2012).
However, other parameters also play a role in triggering this transition.
Young, low surface gravity brown dwarfs retain their clouds to lower {\teff}s than field objects
(Metchev \& Hillenbrand 2006; Saumon \& Marley 2008), as recently demonstrated by the unusually ``dusty'' exoplanets around HR~8799 (Currie et al.\ 2011) and 2MASS~J1207-3932 (Barman et al.\ 2011).  Theoretical work also supports a strong gravity dependence on the onset of the L/T transition (Marley et al.\ 2012). Condensate production appears to be suppressed in metal-poor L dwarfs based on their enhanced TiO and metal-line absorption and extremely blue near-infrared colors (Burgasser et al.\ 2003; Gizis \& Harvin 2006); the transition to the as-yet undiscovered T subdwarf class may be far less dramatic as a result (Burrows et al.\ 2006).
Finally, rotation may play a major role in driving surface inhomogeneities that disrupt cloud layers, although rotation measures (variability periods, $v\sin{i}$) are challenging for these intrinsically faint objects.  Notably, the variable early-T dwarfs SIMP J0136+0933 and 2MASS~J2139+0220, both of which may have highly disrupted cloud decks, also have very different variability periods (2.4~hr and 7.8~hr, respectively).  Additional examples are needed to quantify these effects.

\section{Future work}

The mechanism that drives the L/T transition remains an outstanding problem in brown dwarf atmospheric science, and can be addressed by expanding our sample
and characterization of single and multiple sources spanning this transition.  
Multi-wavelength variability studies hold the promise of confirming or refuting the patchy cloud model for this transition, but proper interpretation of observed lightcurves and the incidence of variability  requires more detailed theoretical work on the 3D structure and temporal variation of rotating brown dwarf atmospheres.  While 2D dynamic models have found that gravity waves can play an important role in stirring the photosphere (Freytag et al.\ 2010), a complete picture will require a 3D calculation that includes rotation effects.  Given the advances in 3D modeling in interpreting brightness variations for transiting exoplanets (Knutson et al.\ 2007) and magnetic topologies of M dwarfs (Browning 2008), such work should be feasible.

Detailed characterization---resolved spectroscopy, parallaxes, and precise orbit determinations---of binaries straddling the L/T transition can provide insight into the role that secondary parameters play in triggering the depletion of condensates; tight spectral binaries in particular can bring mass and age information to this problem.
Finally, there is increasing evidence that sulfide cloud layers are emerging at the transition between the T and Y spectral classes (Morley et al.\ 2012).
Lessons learned from mineral cloud behavior at the L/T transition will be essential for understanding the properties of these very cool atmospheres.

\acknowledgements
The author thanks Mark Marley, Jacqueline Radigan and Didier Saumon for helpful comments in the preparation of this article.

%\bibliographystyle{aa}
%\bibliography{/Users/adam/papers/biblibrary.bib}

\end{document}